# The comparative defect study on the polymeric transfer of MoS$_2$ monolayers

C. Abinash Bhuyan,* Kishore K. Madapu, and Sandip Dhara*

*Email: bhuyan.2d@gmail.com, dhara@igcar.gov.in

Surface and Nanoscience Division, Indira Gandhi Centre for Atomic Research, Homi Bhabha National Institute,

Kalpakkam-603 102, India

Abstract:

The defect-free transfer of chemical vapour deposition (CVD) grown monolayer MoS$_2$ is important for both fabrication of 2D devices and fundamental point of view for various studies where substrate effects need to be minimized. Among many transfer techniques, two well-known techniques that use the polymer as carriers are wet-transfer technique and the surface-energy-assisted transfer technique. In this work, we transferred a single CVD grown monolayer MoS$_2$ by these two transfer methods on a similar substrate, and the intervention of strain and defects in the transfer process is probed by Raman and photoluminescence (PL) spectroscopy, respectively. We found that the conventional and commonly used wet transfer technique degraded the monolayer due to KOH contamination. In contrast, monolayers transferred using the surface-energy-assisted transfer method possess structural integrity and optical quality on a par with the as-grown MoS$_2$ layers. As compared to the wet process a strain-free transfer was recorded in the surface-energy-assisted technique using Raman spectroscopic studies.

**Introduction**

Two-dimensional (2D) transition metal dichalcogenides (TMDs) monolayers are known for their exotic electronic properties and optoelectronic applications.[1-4] In synthesis of high optical quality and large-area monolayer $MoS_2$ (1L-$MoS_2$), chemical vapour deposition (CVD) is proven best among other synthesis methods.[5] However, the high growth temperature and chemical environment of CVD make it challenging to synthesize the monolayers on the polymeric substrates.[6] To overcome this problem, the transfer onto a suitable substrate is important for fundamental and applied research. However, in the transfer process, the inevitable intervention of defects degrades the optical quality of monolayer. Transferring a defect-free monolayer without compromising its optical quality is one of the holy grail in recent 2D material research. Additionally, the multitude of complexity in the CVD synthesised films is so diverse due to growth temperature, reaction time, carrier gases, flow rate and substrate pre-treatment. The fundamental research like defects evaluation in the transfer process was mostly studied on monolayers which were transferred by wet transfer method using poly(methyl methacrylate) (PMMA).[7] A similar method called surface-energy-assisted technique also exists for monolayer transfer which uses polystyrene polymer (PS) as carrier polymer.[8] Till date, the transfer of monolayers grown by CVD technique is carried out in different approaches. Thus, one cannot generalise the merits of the transfer method based on the existing reports.[9,10] In this context, a comparative study of transfer induced defect evaluation lacks in the literature. It is essential to establish the merits of the transfer method with monolayers grown in the same deposition conditions.

Usually, the monolayers are transferred onto flexible substrates such as polyethylene terephthalate (PET) and defect analysis carried out using the photoluminescence (PL) spectroscopy. The defect analysis carried out on these substrates is inhibited by weak emissions from PET substrate which happens to be in the B exciton energy region.[11] So, the evaluation of defects stemmed from the transfer process was manifested using monolayers transferred onto the PET substrate.[4,12] To assess the transfer induced defects, one has to transfer the monolayer onto the same substrate on which it is grown.

The present report evaluates defects in the transferred monolayers which are originated truly because of the transfer process. Here, defect analysis was carried out on two monolayer samples which were transferred by two different approaches. High optical quality $MoS_2$ monolayers were grown on $SiO_2$(300 nm)/Si by atmospheric pressure CVD. To avoid the ambiguity of the substrate effect, we carried out the transfer of monolayer $MoS_2$ onto the $SiO_2$/Si substrate by two transfer methods. The defect analysis was carried out using the Raman and PL spectroscopy.

**Experimental Methods**

*Synthesis of 1L-MoS$_2$*

1L-MoS$_2$ samples were synthesized using the atmospheric pressure chemical vapour deposition technique. Molybdenum trioxide (MoO$_3$, 99.97% Sigma Aldrich) and S ( ≥99.5% pure; Sigma Aldrich) powders were used as precursor materials. Details of the growth process may find elsewhere.[13] The source materials were loaded in a three-zone furnace having one-inch quartz tube as a reaction chamber. The ultra high pure (UHP) Ar gas was used as the carrier gas and the flow rate of carrier gas was controlled by mass flow controller. The polished side of SiO$_2$(300nm)/Si is kept face down over the MoO$_3$ powder in alumina boat. The optimised amount of S (40 mg) and MoO$_3$ (15 mg) were kept in the first and third zone, respectively. Before starting the growth, the chamber was evacuated to a base pressure of $1 \times 10^{-3}$ mbar and purged UHP Ar with the flow rate of 100 sccm for 15 min. The set temperatures of the first, second and third zones were 160, 200 and 700 °C, respectively. However, the ramp rate of three zones was adjusted such that reaching time was the same for all zones. The ramping time of all the zones was 40 mins. The optimised growth time and flow rates were 20 min and 50 sccm, respectively. The furnace was allowed to cool to room temperature naturally after the growth.

*Transfer of 1L-MoS$_2$ onto SiO$_2$(300 nm)/Si substrate*

The transfer of 1L-MoS$_2$ was carried out in two different methods. The difference in these methods was the carrier polymer such as PMMA and PS. The steps of each approach are explained in the following section.

**Method-1:** The conventional wet transfer technique is previously reported.[7] However, we briefly outline the steps in the transfer process. A 100 mg of PMMA crystals (MW-80,000 g/mol, Sigma Aldrich) were mixed with 1 ml of anisole (Merck) to prepare the PMMA solution. Subsequently, the prepared PMMA solution was spin-coated over the 1L-MoS$_2$/SiO$_2$(300 nm)/Si substrate with 4000 RPM for 30 sec and subsequently, the substrate was baked for 30 min at 60 °C. In the next step, the polymer capped PMMA/1L-MoS$_2$/SiO$_2$(300 nm)/Si substrate was dipped in KOH (1M) solution for 30 min. The bubbles formed on the SiO$_2$ substrate helped to lift-off the PMMA capped monolayer. Then, PMMA capped 1L-MoS$_2$ was fished-out by another cleaned SiO$_2$/Si substrate and which was followed by baking at 90 °C for 30 min for good adhesion. After the baking process, the removal of PMMA films was carried out by rinsing with the acetone several times. Then, the transferred film was annealed at 100 °C under UHP Ar atmosphere for 2 hr.

**Method-2:** In another approach, we adopted the surface-energy-assisted transfer method to transfer 1L-MoS$_2$ onto SiO$_2$/Si.[8] In this process, 180 mg of polystyrene crystals (MW-280, 000 g/mol, Sigma Aldrich) was mixed in 2 ml toluene (Merck) to form the uniform PS solution. Then, the PS solution was spin-coated on 1L-MoS$_2$/SiO$_2$(300 nm)/Si substrate by 3500 RPM for 60 s. Subsequently, the PS capped monolayer was baked at 90 °C for 15 min for good adhesion. After the baking process, PS capped monolayer was dipped in the distilled water and the water was allowed to enter between PS capped monolayer and SiO$_2$/Si substrate by poking a sharp needle near the edge region. Because of the hydrophobic nature of PS, water penetrated between the SiO$_2$ and PS which subsequently helped in lift-off the 1L-MoS$_2$ capped with PS. The floating 1L-MoS$_2$ capped with PS film was fished-out onto a new cleaned SiO$_2$ substrate. In the next step, the SiO$_2$ substrate with PS capped 1L-MoS$_2$ was baked at 100 °C for 5 min. Subsequently, the PS was removed by dissolving it in toluene. Then, the transferred film was annealed in 100 °C under UHP Ar atmosphere for 2 hr.

**Optical spectroscopic characterization:**

Raman and PL analysis was carried out using the micro-Raman spectrometer (inVia, Renishaw, UK). The wavelength of excitation was 532 nm and scattered light was collected in backscattering geometry. All Raman and PL spectra were collected by a thermoelectrically cooled charge-coupled device detector after dispersing through 2400 gr/mm and 1800 gr/mm gratings, respectively.

**Result and discussion:**

The growth of 1L-MoS$_2$ is confirmed by optical microscopy. Fig. 1a shows the typical optical micrograph of the as-grown 1L-MoS$_2$ flakes on SiO$_2$/Si substrate. The as-grown monolayers were transferred onto fresh SiO$_2$/Si by two methods which were discussed in the experimental sections. Figures 1b and 1c show the optical images of transferred 1L-MoS$_2$ flakes by method-1 and method-2, respectively and illustrate the damage- and wrinkle-free monolayer domains. The 2D TMDs are well characterised by Raman spectroscopy owing to the covalent bonding between the constituent atoms.[14] The 1L-MoS$_2$ shows the two distinct Raman modes which are assigned as $E^1_{2g}$ and $A_{1g}$ phonon modes and these modes correspond to the in-plane vibrations of S and Mo atoms and out-of-plane vibrations of S atoms,

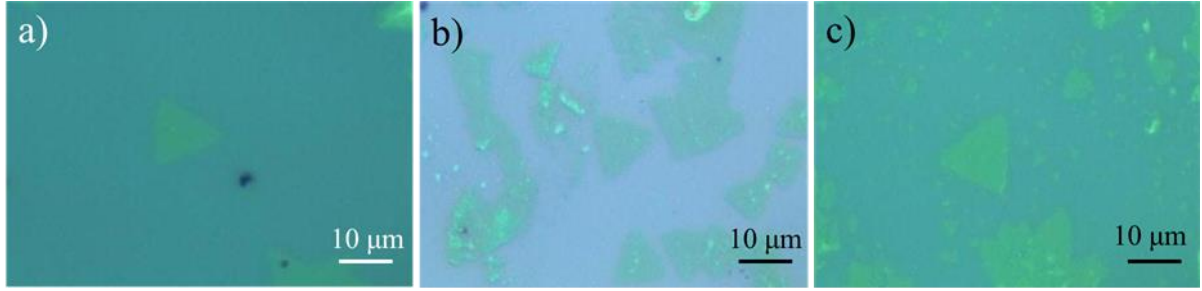

Figure 1 (a-c) The optical micrographs of as-grown, transferred 1L-MoS$_2$ triangular flakes by method-1 and method-2 respectively.

respectively. Additionally, longitudinal optical LO(M) and longitudinal acoustic 2LA(M) modes are also usually observed in monolayer Raman spectra.[15] Figures 2a-c show the typical Raman spectra of as-grown and transferred 1L-MoS$_2$ flakes by method 1 and method 2, respectively. One can assure the structural integrity of transferred 1L-MoS$_2$ flakes by using the Raman spectra (Figs. 2b and 2c). In our as-grown monolayer, the $E^1_{2g}$ and $A_{1g}$ mode frequencies were found to be 386.16 and 404.77 cm$^{-1}$, respectively. The difference in the frequency of Raman modes ($\Delta$) is 19 cm$^{-1}$ which confirms that monolayer thickness of the MoS$_2$ flake.[15] However, 1L-MoS$_2$ transferred by method-1 shows red-shift in $E^1_{2g}$ mode by 0.88 cm$^{-1}$. In contrast, the $A_{1g}$ mode is blue-shifted by 1.27 cm$^{-1}$. The difference between two modes show a value of 20.7 cm$^{-1}$ which further confirms the structural integrity of the transferred monolayer (Fig. 2b). This opposite behaviour can be understood by the in-plane compressive strain and stiffening. Usually, the $E^1_{2g}$ mode is influenced by in-plane compressive strain.[13] As reported previously, the shift in $E^1_{2g}$ mode is 2.1 cm$^{-1}$ per % of uniaxial strain,[16] and 4.7 cm$^{-1}$ per % for biaxial strain.[17] Considering these values, we calculated the uniaxial and biaxial strain in our transferred monolayer as ~ 0.41% and 0.18%, respectively (Fig. 2b). Thus, the observed red-shift is attributed to the relaxation of compressive strain because of weak adhesion of transferred film as compared to the as-grown film. However, the contrasting behaviour of $A_{1g}$ mode is attributed to the stiffening of phonon mode because of doping or residual polymer molecules perturbing the vibrational potential landscape of the transferred film.[18,19] The monolayer transferred by method-2 shows Raman modes centred at 386.17 and 406.38 cm$^{-1}$ for the $E^1_{2g}$ and $A_{1g}$ phonon modes, respectively (Fig. 2c). In this case also the $A_{1g}$ mode was blue-shifted as compared to as-grown 1L-MoS$_2$ and the same reason is applicable here. There was no measurable shift in the $E^1_{2g}$ mode indicating a strain-free transfer process. Only Raman measurement is insufficient to differentiate the doping or residual polymer molecules as the primary cause of both the transferred film. So, more insights were obtained by employing the PL spectroscopy.

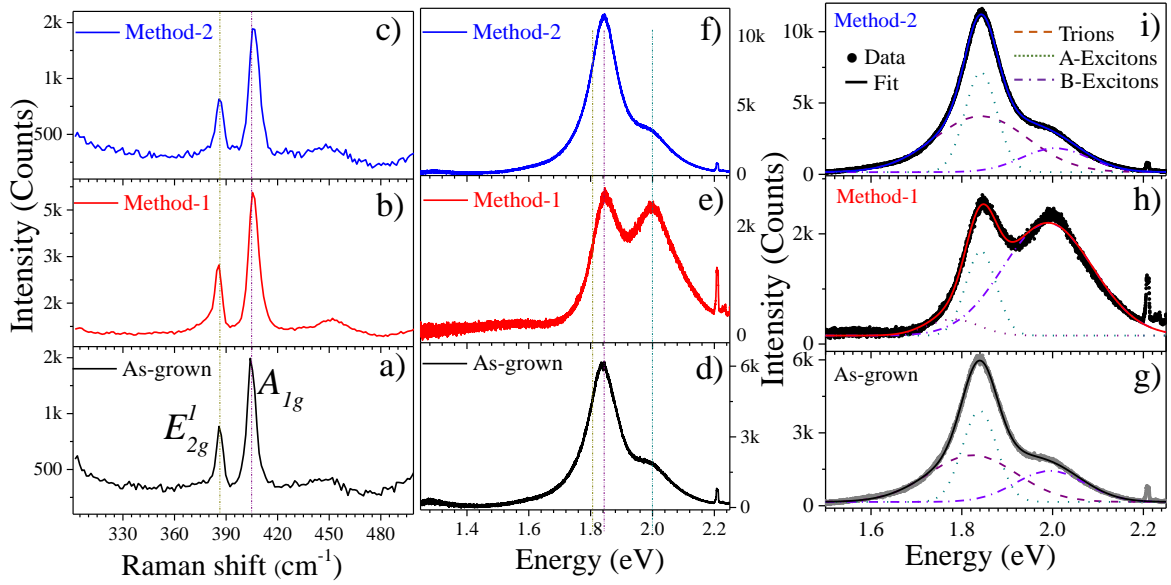

Figure 2 (a-c) The ambient Raman spectra of as-grown, transferred monolayer by method-1 and method-2 respectively. (d-f) Ambient PL spectra of as-grown, transferred monolayer by method-1 and method-2 respectively. (g-i) Gaussian curve fitting of respective PL spectra showing trions (~1.82 eV), A-exciton (~1.85 eV) and B-exciton (~1.99 eV).

To further probe the optical quality and intrinsic material properties of transferred 1L-MoS$_2$, PL spectroscopy is employed. As shown in figures 2d-f and in the fitted data (Figs 2g-i), the ambient PL spectra of 1L-MoS$_2$ have the major contribution from A-exciton emission at ~1.85 eV and minor contributions from trions (~1.82 eV) and B-excitons (~1.99 eV).[20] Also, the intensity ratio of A- to B-exciton ($I_A/I_B$) measures the optical quality of monolayer TMDCs.[21] The monolayer transferred by method-1 shows the drastic decrease in A-exciton intensity in comparison to B-exciton intensity. Whereas, B-exciton intensity is almost unchanged after the transfer. The decreasing $I_A/I_B$ value indicates the increase in defect density for transferred 1L-MoS$_2$.[21] The $I_A/I_B$ for as-grown and transferred 1L-MoS$_2$ by methods-1 were 2.72 and 0.76, respectively (Figs. 2g and 2h). In figure 2h, the A- and B- exciton shows almost comparable peak intensities which manifests the non-radiative recombination because of the creation of a high amount of defects.[21] Employing that argument, we can conclude that defects density is high in the case of monolayers transferred by method-1. The negligible blue-shift value 10 meV for A-exciton is attributed to the compressive strain in the transferred film originated from residual polymer molecules. In contrast, the film transferred by method-2 shows the increase of individual exciton intensities and $I_A/I_B$ value to 3.88 (Fig. 2i). The high $I_A/I_B$ value indicates low defect density and high sample quality. Thus, it can be understood that the quality of the monolayer is retained after the transfer by method-2. Moreover, the total PL intensity is more than as-grown 1L-MoS$_2$ PL intensity. This behaviour may have originated because of the optical interference effect due to multiple reflections from the SiO$_2$ substrates. This effect is most prominent in transferred flakes because of the substrate is clean and free from any S stacking or precursor intermediates as compared to the grown substrates.[22] Nevertheless, the optical quality of the monolayer film was maintained in 1L-$MoS_2$ transferred by method-2. The peak position A-exciton is blue-shifted (10 meV) as compared to the as-grown films because of compressive strain originated from the residual polymer molecules as also observed in the case of method-1 (Fig. 2e). In nutshell, the sample transferred by PS as carrier polymer (method-2) is less prone to defects in the transfer process as compare to transfer by PMMA as carrier polymer (method-1). Moreover, the transfer of the film by surface-energy-assisted method onto the polymer substrate may also improve the efficiency of flexible 2D devices.

**Conclusion**

In the present study, we find that the conventional wet transfer technique introduces more defects in the transfer process as compared to the surface-energy-assisted method as studied using photoluminescence studies at ambient. The overall reduction in photoluminescence intensity by former transfer method provides non-radiative pathways for exciton annihilation which is due to the transfer-induced defects. The former transfer process also shows a uniaxial and biaxial strain of ~ 0.41% and 0.18%, respectively, in the transferred monolayer. Whereas a strain-free transfer process is realized in the latter method. This finding helps the transfer of high-quality monolayer onto different substrates. Thus surface-energy-assisted transfer process may be extended to other 2D materials for studying defect evaluation and for the further fabrication of efficient 2D optoelectronic devices.

**Acknowledgements**

We thank Mr. Gopinath Sahoo for his valuable suggestions during film transfer work.

**Conflicts of interest**

There are no conflicts of interest to declare.